\documentclass{iopart}
\usepackage{iopams}
\usepackage{geometry}                
\usepackage{graphicx}
\usepackage{epstopdf}
\usepackage{amsthm}

\usepackage{cite}

\usepackage{url}

\usepackage[percent]{overpic}

\usepackage{color}

\begin{document}

\title{A novel 8-parameter integrable map in $\mathbb{R}^4$}

\author{G R W Quispel, D I McLaren and P H van der Kamp}

\address{Department of Mathematics,
	La Trobe University,
	Bundoora, VIC 3083, Australia   \ead{r.quispel@latrobe.edu.au}\ead{d.mclaren@latrobe.edu.au}\ead{p.vanderkamp@latrobe.edu.au}}

\begin{abstract}
We present a novel 8-parameter integrable map in $\mathbb{R}^4$. The map is measure-preserving and possesses two functionally independent 2-integrals, as well as a measure-preserving 2-symmetry.
\end{abstract}

\section{Introduction}
\label{intro}
Discrete integrable systems have attracted a lot of attention in recent years \cite{HJN}. One of the reasons for this comes from physics: many physical models include discreteness at a fundamental level. Another reason for the increased interest in discrete integrable systems comes from mathematics: in several instances it turns out that discrete integrable systems are arguably richer, or more fundamental than continuous (i.e. non-discrete) ones. Prime examples are (i) Integrable partial difference equations (P$\Delta$Es), where a single P$\Delta$E yields (through the use of vertex operators) an entire infinite hierarchy of integrable partial differential equations \cite{WC}; (ii) Discrete Painlev\'{e} equations, where the Sakai classification is much richer in the discrete case than in the continuous one \cite{S}; (iii) Darboux polynomials, where in the discrete case unique factorization of the so-called co-factors can be used (which does not exist in the continuous (additive) case)\footnote{cf \cite{CEMOQT,CEMOQTV} for the discrete case, and \cite{Goriely} for a very nice introduction to the continuous case. }.

\medskip \noindent In this Letter we will be interested in autonomous integrable ordinary difference equations (or maps). Much interest was generated by the discovery of the 18-parameter integrable QRT map in $\mathbb{R}^2$ (\cite{JJD,QRT1,QRT2}). For some other examples in higher dimensions, cf. e.g. Chapter 6 of \cite{HJN}.

\medskip \noindent A special aspect of the maps we consider in this Letter is that they are an example of integrable maps arising as discretisations of ordinary differential equations (ODEs). Earlier examples of this arose using the Kahan discretisation of first-order quadratic ODEs (cf. \cite{Kahan}, \cite{HK}, \cite{PPS}, \cite{CMMOQ2} and references therein), or by the discretisation of ODEs of order 1 and arbitrary degree using polarisation methods \cite{CMMOQ1}, and by the methods in \cite{HQ} for the discretisation of ODEs of order $o$ and degree $o+1$, cf. also \cite{M,McL}.

\medskip \noindent In section 3 we present a novel integrable 8-parameter map in $\mathbb{R}^4$. This map generalizes a 5-parameter map in $\mathbb{R}^4$ found earlier in \cite{CMMOQ1} to the inhomogeneous case, and because the derivation of the novel map may be somewhat mysterious if the reader is unfamiliar with the previous map and its derivation, we summarise the latter in section 2.

\section{What went before}
\label{sec:1}
In \cite{CMMOQ1} Celledoni, McLachlan, McLaren, Owren and Quispel introduced a novel integrable map in $\mathbb{R}^4$. It was constructed as follows.

\medskip \noindent The authors considered the homogeneous quartic Hamiltonian
\begin{equation}\label{homQham}
H = aq^4 + 4bq^3p + 6cq^2p^2 + 4dqp^3 + ep^4,
\end{equation}
where $a,b,c,d$ and $e$ are 5 arbitrary parameters.

\medskip \noindent This gave rise to an ordinary differential equation (ODE)
\begin{equation}\label{ode3}
\frac{d}{dt} \left( \begin{array}{c} q \\ p \end{array} \right)
= \left(\begin{array}{rr} 0 & 1 \\ -1 & 0 \end{array}  \right)  \nabla H = f_3  \left( \begin{array}{c} q \\ p \end{array} \right) ,
\end{equation}
where the cubic vector field $f_3$ is defined by
\begin{equation}\label{f3}
f_3  \left( \begin{array}{c} q \\ p \end{array} \right)  =
\left( \begin{array}{c} 4bq^3 +12cq^2p + 12dqp^2 + 4ep^3    \\
 -4aq^3 - 12bq^2p - 12 cqp^2 - 4dp^3 \end{array} \right).
\end{equation}
Defining $x:=\left( \begin{array}{c} q \\ p \end{array} \right)$, and introducing the timestep $h$, the vector field (\ref{ode3}) was then discretized:
\begin{equation}\label{map3}
\frac{x_{n+2} - x_n}{2h} = F_3(x_n,x_{n+1},x_{n+2}),
\end{equation}
where $F_3$ was defined using polarization, i.e.
\begin{equation}\label{poldef}
F_3(x_n,x_{n+1},x_{n+2}) := \frac{1}{6}  \frac{\partial}{\partial \alpha_1} \frac{\partial}{\partial \alpha_2} \frac{\partial}{\partial \alpha_3} f_3(\alpha_1x_n + \alpha_2x_{n+1} + \alpha_3x_{n+2}) |_{\alpha=0} 
\end{equation}
It is not difficult to check that the multilinear function $F_3$ defined by (\ref{poldef}) is equivalent to
\begin{eqnarray}
F_3(x_n,x_{n+1},x_{n+2}) &:=&  \frac{9}{2}f_3 \left( \frac{x_n + x_{n+1} + x_{n+2}}{3} \right) - \frac{4}{3} f_3\left( \frac{x_n + x_{n+1}}{2} \right)\nonumber \\
&-& \frac{4}{3} f_3\left( \frac{x_n + x_{n+2}}{2} \right) - \frac{4}{3} f_3\left( \frac{x_{n+1} + x_{n+2}}{2} \right) \nonumber \\
&+& \frac{1}{6} f_3\left( x_n \right) + \frac{1}{6} f_3\left( x_{n+1} \right) + \frac{1}{6} f_3\left( x_{n+2} \right),
\end{eqnarray}
cf \cite{CMMOQ1} and page 110 of reference \cite{Greenberg}.

\medskip \noindent By construction, the rhs of (5) is linear in $x_{n+2}$ and $x_n$ for cubic vector fields, i.e. (\ref{map3}) represents a birational map (see \cite{CMMOQ1}), and it was shown that this map possesses two functionally independent 2-integrals (recall that a 2-integral of a map $\phi$ is defined to be an integral of $\phi \circ \phi$):
\begin{eqnarray}
I(q_n,p_n,q_{n+1},p_{n+1}) &=& q_n p_{n+1} - p_n q_{n+1}  \\
I(q_{n+1},p_{n+1},q_{n+2},p_{n+2}) &=& q_{n+1} p_{n+2} - p_{n+1} q_{n+2} ,
\end{eqnarray}
where $q_{n+2}$ and $p_{n+2}$ should be eliminated from (7) using (\ref{map3}).

\medskip \noindent Note that (7) above does not depend on the parameters $a,b,c,d,e$ (in contrast to (8), which will depend on the parameters once expressed in $q_n,q_{n+1},p_n,p_{n+1}$).

\medskip \noindent The map (\ref{map3}) also preserves the measure
\begin{equation}\label{meas4}
\frac{dq_n \wedge dp_n \wedge dq_{n+1} \wedge dp_{n+1}}{1 + 4h^2 \Delta_1},
\end{equation}
where\footnote{Erratum: In eqs (4.1) of \cite{CMMOQ1}, $1-4h^2\Delta$ should read $1+4h^2\Delta$. Their $\Delta$ is our $\Delta_1$.} 
\begin{eqnarray}\label{Delta4}
\Delta_1 &=& \left| \begin{array}{cc} c & d \\ d & e \end{array}  \right| p_n^2p_{n+1}^2 + 
\left| \begin{array}{cc} b & c \\ d & e \end{array}  \right| (p_n^2p_{n+1}q_{n+1} + p_nq_nq_{n+1}^2) \\
&+& \left| \begin{array}{cc} b & c \\ c & d \end{array}  \right| (p_n^2q_{n+1}^2 + q_n^2p_{n+1}^2) +
 \left| \begin{array}{cc} a & b \\ c & d \end{array}  \right| (q_n^2p_{n+1}q_{n+1} + p_nq_nq_{n+1}^2) \nonumber \\ 
&+&\left| \begin{array}{cc} a & c \\ c & e \end{array}  \right| p_nq_np_{n+1}q_{n+1} + 
\left| \begin{array}{cc} a & b \\ b & c \end{array}  \right| q_n^2q_{n+1}^2. \nonumber
\end{eqnarray}

\medskip \noindent Finally, the map (\ref{map3}) is invariant under the scaling symmetry group
\begin{equation}\label{scale}
x_n \rightarrow  \lambda^{(-1)^n}x_n.
\end{equation}

\section{A novel 8-parameter integrable map in $\mathbb{R}^4$}
We now generalise the treatment of section 2 to the non-homogeneous Hamiltonian
\begin{equation}\label{inhomQham}
H = aq^4 + 4bq^3p + 6cq^2p^2 + 4dqp^3 + ep^4 + \frac{1}{2}\rho q^2 + \sigma qp + \frac{1}{2} \tau p^2,
\end{equation}
where $a,b,c,d,e,\rho,\sigma$ and $\tau$ are 8 arbitrary parameters.

\medskip \noindent This gives rise to an ODE
\begin{equation}\label{ode31}
\frac{d}{dt} \left( \begin{array}{c} q \\ p \end{array} \right)
=\left(\begin{array}{rr} 0 & 1 \\ -1 & 0 \end{array}  \right)  \nabla H =  f_3  \left( \begin{array}{c} q \\ p \end{array} \right)  + f_1  \left( \begin{array}{c} q \\ p \end{array} \right),
\end{equation}
where the cubic part of the vector field, $f_3$, is again given by (\ref{f3}), whereas the linear part $f_1$ is given by
\begin{equation}\label{f1}
f_1  \left( \begin{array}{c} q \\ p \end{array} \right)  =
\left( \begin{array}{c} \sigma q + \tau p    \\
 -\rho q - \sigma p \end{array} \right).
\end{equation}
We now discretise the cubic part resp. the linear part of the vector field in different ways:
\begin{equation}\label{map31}
\frac{x_{n+2} - x_n}{2h} = F_3(x_n,x_{n+1},x_{n+2}) + F_1(x_n,x_{n+2}),
\end{equation}
where $F_3$ is again defined by (5), but $F_1$ is defined by a kind of midpoint rule:
\begin{equation}\label{F1}
F_1(x_n,x_{n+2}) = f_1 \left( \frac{x_n + x_{n+2}}{2} \right).
\end{equation}
It follows that equation (\ref{map31}) again defines a birational map, and, importantly, it again preserves the scaling symmetry (\ref{scale}). (Indeed the latter is the primary reason we use the discretization (\ref{F1})).

\medskip \noindent Two questions thus remain:
\begin{enumerate}
\item Does eq (\ref{map31}) preserve two 2-integrals?
\item Is eq (\ref{map31}) measure-preserving?
\end{enumerate}
The answer to both these questions will turn out to be positive.

\medskip \noindent We actually had numerical evidence several years ago that the map (\ref{map31}) (or at least a special case of it) was integrable. However it has taken us until now to actually find closed-form expressions for the preserved measure and for the 2-integrals.

\medskip \noindent A first clue to the identity of a possible 2-integral of (\ref{map31}) came when we were carrying out experimental mathematical computations (in the sense of \cite{BB}) to find ``discrete Darboux polynomials'' for the map (\ref{map31}) (cf. \cite{CEMOQT} and \cite{CEMOQTV}). This gave a hint that a possible quadratic 2-integral $I(q_n,p_n,q_{n+1},p_{n+1})$ generalising (7), might exist for the map (\ref{map31}).

\medskip \noindent However, the mathematical complexity of the general 8-parameter map (\ref{map31}) was too great  to carry out these computations for a completely general quadratic 2-integral in four variables with all 8 parameters symbolic. 

\medskip \noindent Our process of discovery thus proceeded in two steps:

\medskip \noindent Step 1: Taking all parameters $a,b,c,d,e,\rho,\sigma,\tau$ and $h$ to be random integers, and assuming the 2-integral was an arbitrary quadratic function in four variables (with coefficients to be determined), we computed the 2-integral for a large number of random choices of the integer parameters. In each case, it turned out that the same six coefficients in the quadratic function were zero, i.e. the 2-integral always had the form
\begin{equation}\label{2int1}
I(q_n,p_n,q_{n+1},p_{n+1}) = A q_nq_{n+1} + B p_np_{n+1} + C q_np_{n+1} + D p_nq_{n+1},
\end{equation}
where $A,B,C$, and $D$ depended on the parameters in a way as yet to be determined.

\medskip \noindent Step 2: Now taking all parameters $a,b,c,d,e,\rho,\sigma,\tau$ and $h$ symbolic, and assuming the 2-integral $I$ had the special quadratic form (\ref{2int1}), we found
\begin{eqnarray}\label{2int2}
I(q_n,p_n,q_{n+1},p_{n+1}) = (h\sigma +1)p_nq_{n+1} &+& (h\sigma -1) q_n p_{n+1}  \\
& &+ h\rho q_nq_{n+1} + h\tau p_np_{n+1}. \nonumber
\end{eqnarray}

\medskip \noindent Notes:

\begin{enumerate}
\item The 2-integral (\ref{2int2}) is invariant under the scaling symmetry group (\ref{scale})\footnote{The scaling symmetry (\ref{scale}) is an essential ingredient in our proof of the Theorem in the current Letter that the map (\ref{map31}) is integrable (as well as in our proof in \cite{CMMOQ1} that the map (\ref{map3}) is integrable).}.
\item In the continuum limit $h \rightarrow 0$, and using eq (\ref{ode31}), the integral $I(q_n,p_n,q_{n+1},p_{n+1})/h \rightarrow 4H(q,p)$.
\item Like equation (7), equation (\ref{2int2}) does not explicitly depend on the parameters $a,b,c,d,e$.
\item Note that it is a common feature of many dynamical systems that one has a choice to either study a given phenomenon for a single system containing as many free parameters as possible, or alternatively for multiple systems in so-called normal form (obtained by suitable transformations of the variables), containing fewer parameters. Both in our earlier works on the QRT map \cite{QRT1,QRT2}, and on the 5-parameter map in $\mathbb{R}^4$ \cite{CMMOQ1}, as well as in the current Letter, we have chosen the former option.
\end{enumerate}

\medskip \noindent Once we had the putative equation (\ref{2int2}), it was not difficult to verify using symbolic computation that $I(q_n,p_n,q_{n+1},p_{n+1})$ and $I(q_{n+1},p_{n+1},q_{n+2},p_{n+2})$ are indeed functionally independent 2-integrals of (\ref{map31}).

\medskip \noindent The map (\ref{map31}) preserves the measure
\begin{equation}\label{meas42}
\frac{dq_n \wedge dp_n \wedge dq_{n+1} \wedge dp_{n+1}}{1 + 4h^2 (\Delta_1 + \Delta_2)},
\end{equation}
where the quartic function $\Delta_1$ is given by (\ref{Delta4}) and the quadratic function $\Delta_2$ is given by
\begin{eqnarray}\label{Delta42}
\Delta_2 &=& \frac{1}{2} \left( \left| \begin{array}{cc} a & b \\ \sigma & \tau \end{array}  \right| + \left| \begin{array}{cc} c & b \\ \sigma & \rho \end{array}  \right| \right) q_nq_{n+1} +
\frac{1}{2} \left( \left| \begin{array}{cc} c & d \\ \sigma & \tau \end{array}  \right| + \left| \begin{array}{cc} e & d \\ \sigma & \rho \end{array}  \right| \right) p_np_{n+1} \\
&+& \frac{1}{2} \left( \left| \begin{array}{cc} b & c \\ \sigma & \tau \end{array}  \right| + \left| \begin{array}{cc} d & c \\ \sigma & \rho \end{array}  \right| \right) (p_nq_{n+1} + q_np_{n+1}) + \frac{1}{4} \left| \begin{array}{cc} \rho & \sigma \\ \sigma & \tau \end{array}  \right| .\nonumber
\end{eqnarray}

\medskip \noindent Finally, the map (\ref{map31}) is again invariant under the scaling symmetry group (\ref{scale}).

\medskip \noindent {\bf Theorem} The birational map defined by (\ref{map31}) is integrable.

\medskip \noindent {\it Proof} The proof of integrability is identical to the proof in \cite{CMMOQ1}. The second iterate of the map defined by (\ref{map31}) has a one-dimensional measure-preserving symmetry group. The map thus descends to a measure-preserving map on the three-dimensional quotient. The two integrals of the second iterate of the map are invariant under the symmetry and therefore also pass to the quotient. This yields a three-dimensional measure-preserving map with two integrals, which is thus integrable.

\medskip \noindent {\bf Acknowledgements}

\medskip \noindent We are grateful to R. McLachlan for early discussions on scaling symmetry, and to E. Celledoni, B. Owren and B. Tapley for many discussions on discrete Darboux polynomials.

\end{document}